\documentclass[article]{jss}\usepackage[]{graphicx}\usepackage[]{xcolor}
\makeatletter
\def\maxwidth{ %
  \ifdim\Gin@nat@width>\linewidth
    \linewidth
  \else
    \Gin@nat@width
  \fi
}
\makeatother

\definecolor{fgcolor}{rgb}{0.345, 0.345, 0.345}
\newcommand{\hlnum}[1]{\textcolor[rgb]{0.686,0.059,0.569}{#1}}%
\newcommand{\hlsng}[1]{\textcolor[rgb]{0.192,0.494,0.8}{#1}}%
\newcommand{\hlcom}[1]{\textcolor[rgb]{0.678,0.584,0.686}{\textit{#1}}}%
\newcommand{\hlopt}[1]{\textcolor[rgb]{0,0,0}{#1}}%
\newcommand{\hldef}[1]{\textcolor[rgb]{0.345,0.345,0.345}{#1}}%
\newcommand{\hlkwa}[1]{\textcolor[rgb]{0.161,0.373,0.58}{\textbf{#1}}}%
\newcommand{\hlkwb}[1]{\textcolor[rgb]{0.69,0.353,0.396}{#1}}%
\newcommand{\hlkwc}[1]{\textcolor[rgb]{0.333,0.667,0.333}{#1}}%
\newcommand{\hlkwd}[1]{\textcolor[rgb]{0.737,0.353,0.396}{\textbf{#1}}}%

\usepackage{framed}
\makeatletter
\newenvironment{kframe}{%
 \def\at@end@of@kframe{}%
 \ifinner\ifhmode%
  \def\at@end@of@kframe{\end{minipage}}%
  \begin{minipage}{\columnwidth}%
 \fi\fi%
 \def\FrameCommand##1{\hskip\@totalleftmargin \hskip-\fboxsep
 \colorbox{shadecolor}{##1}\hskip-\fboxsep
     \hskip-\linewidth \hskip-\@totalleftmargin \hskip\columnwidth}%
 \MakeFramed {\advance\hsize-\width
   \@totalleftmargin\z@ \linewidth\hsize
   \@setminipage}}%
 {\par\unskip\endMakeFramed%
 \at@end@of@kframe}
\makeatother

\definecolor{shadecolor}{rgb}{.97, .97, .97}
\definecolor{messagecolor}{rgb}{0, 0, 0}
\definecolor{warningcolor}{rgb}{1, 0, 1}
\definecolor{errorcolor}{rgb}{1, 0, 0}
\newenvironment{knitrout}{}{} 

\usepackage{alltt}

\usepackage{orcidlink,lmodern}

\usepackage[american]{babel}
\usepackage{amsmath}
\usepackage{amssymb}%
\usepackage{mathrsfs}
\usepackage{amsbsy}
\usepackage{amsthm}
\usepackage{float}
\usepackage{booktabs}
\usepackage{underscore}

\newcommand{\class}[1]{`\code{#1}'}
\newcommand{\fct}[1]{\code{#1()}}

\graphicspath{{figures/}}

\author{Giovanni Maria Merola~\orcidlink{0000-0003-2539-9225}\\
Independent researcher}
\Plainauthor{Giovanni Maria Merola} 

\title{An R package to Compute Least Squares Sparse Principal Components}
\Plaintitle{An R package to compute Least Squares Sparse Principal Components}
\Shorttitle{An \proglang{R} package to compute LS-SPCA components}

\Abstract{
This paper introduces the \proglang{R} package \pkg{spca}, which provides a computational framework for least squares sparse principal component analysis (LS-SPCA). Unlike other SPCA methods, LS-SPCA generates uncorrelated sparse principal components (sPCs) that effectively maximize the explained variance while maintaining strong correlations with standard principal components (PCs). The framework also includes more computationally efficient variants that produce mildly correlated sPCs, which often have lower cardinality while explaining equal or greater variance than the LS-SPCA optimal sPCs. 

The \pkg{spca} package is built on an efficient \proglang{C++} backend for matrix computations, with distinct engines for tall and fat matrices, and a flexible \proglang{R} frontend. The user interface offers several options for computing sPCs, such as deciding whether sparsification should stop when a threshold for cumulative variance explained or $R^2$ with the PCs is reached, and choosing between simple forward selection, stepwise forward selection, or backward elimination for variable selection. In addition to the \fct{print}, \fct{summary}, and \fct{plot} methods, the package includes tools for comparing different \class{spca} solutions, grouping sparse loadings, and representing foreign SPCA solutions as \class{spca} objects. 

This article demonstrates with real datasets the use of the package in a typical LS-SPCA workflow and briefly contrasts LS-SPCA with conventional SPCA solutions . Then it compares different LS-SPCA solutions obtained from the dataset. Finally, the performance of \pkg{spca} on large tall and fat matrices is discussed, showing that \pkg{spca} offers a computationally efficient alternative for computing interpretable sPCs.
}
\Plainkeywords{SPCA, LS-SPCA, variance explained, projection, uncorrelated, R}

\Address{
  Giovanni Maria Merola\\
  Independent Researcher\\
  Vietnam\\
  E-mail: \email{merolagio@gmail.com}\\
}

\makeatletter
\newcommand\codo{%
  \begingroup
  \@makeother\_\@makeother\~\@makeother\$\@codox
}
\def\@codox#1{%
  {\normalfont\ttfamily\hyphenchar\font=-1 \detokenize{#1}}%
  \endgroup
}
\makeatother

\newcommand{\trasp}{^{\scriptstyle{\intercal}}}
\newcommand{\traspj}{^{\scriptstyle{\intercal}}_j}
\newcommand{\trasps}[1]{^{\raisebox{\depth}{$\scriptstyle{\intercal}$}}_{#1}} 
\newcommand{\traspd}[1]{^{\raisebox{-1.0mm}{$\scriptstyle{\intercal}$}}_{#1}} 

\newcommand{\traspdr}[1]{^{\raisebox{-1.0mm}{$\scriptstyle{\intercal}$}}_{(#1)}}

\newcommand{\QjT}{{{Q}\trasps{j}}}

\newcommand{\daj}{{\overset{\boldsymbol{.}}{a}_j}}

\newcommand{\dA}{{\overset{\boldsymbol{.}}{A}}}

\newcommand{\da}{{\overset{\boldsymbol{.}}{a}}}

\newcommand{\dXj}{{\overset{\boldsymbol{.}}{X}_j}}
\newcommand{\dXjT}{{\overset{\boldsymbol{.}}{{X}}\traspd{j}}}

\newcommand{\dSj}{{\overset{\boldsymbol{.}}{S}_j}}
\newcommand{\dSjT}{{\overset{\boldsymbol{.}}{S}\traspd{j}}}

\newcommand{\matinvj}[1]{^{\raisebox{-1.0mm}{$\scriptstyle{{-1}}$}}_{#1}} 
\IfFileExists{upquote.sty}{\usepackage{upquote}}{}
\begin{document}
\section{Introduction}
Principal component analysis (PCA) is a widely used technique for reducing the dimensionality of multivariate data. PCA achieves this by replacing the original variables with linear combinations, called principal components (PCs), which are arranged in order of decreasing variance explained. Typically, the first few PCs capture a significant portion of the data's variation. This is accomplished by requiring the PCs to optimally approximate the data in the least-squares (LS) sense.

Interpreting PCs becomes challenging when the number of variables is large, as all variables contribute to each component. Moreover, the common practice of focusing solely on 'large' loadings for interpretation often results in the selection of highly correlated variables that collectively possess little greater explanatory power than that of a single variable \citep{mer20}.

Sparse principal component analysis (SPCA) addresses the interpretability issue by computing \emph{sparse} principal components (sPCs). These sPCs are linear combinations of only a few variables, with several coefficients, referred to as \emph{loadings}, being zero. This sparsity simplifies the identification of key variables that define the PCs.

The first SPCA method was introduced by \cite{tre}, and numerous other SPCA methods have been developed since. These \emph{conventional} SPCA methods determine sPCs by applying sparsity constraints to the PCA objective of maximizing the variance of the PCs, as originally proposed by \citet{hot}.

Conventional SPCA methods differ in terms of the sparsity constraints they impose, the transformation of data post-sPC computation, and the optimization techniques employed, thereby offering diverse solutions to the same problem. In some cases, the resulting sPCs deviate significantly from the original PCs and exhibit mutual correlation. The review series \emph{All sparse PCA models are wrong, but some are useful} (\citeauthor{camI}, \citeyear{camI}, \citeyear{camII}, \citeyear{camIII}) provides an analysis of many existing conventional SPCA methods. Additionally, \citet{mer15} and  \citet{mer19} discuss the conventional SPCA approach.

Least squares sparse principal component analysis (LS-SPCA) was initially proposed by \citet{mer15}, with its projection variant introduced by \citet{mer19}. This method yields uncorrelated sPCs that genuinely maximize the explained variance by directly imposing sparsity constraints on Pearson's classic PCA data-approximation objective \citep{pea}.

The difference in results between conventional SPCA and LS-SPCA arises because Pearson's and Hotelling's objectives yield identical solutions only in standard PCA. When sparsity constraints are imposed, the solutions differ \citep{mer15, mer19}. Consequently, LS-SPCA tends to select variables that explain the residual variance of the entire dataset, whereas conventional SPCA identifies sets of variables best explained by their first principal component \citep{mog}.

The \pkg{spca} package efficiently computes LS-SPCA solutions and includes functions for visualizing and comparing these solutions. The primary computations are executed by the \proglang{C++} backend, which is optimized to reduce computational effort, such as employing a reverse-SVD approach for \code{fat} matrices. The package offers three methods for printing, plotting, and summarizing results. Additionally, it includes tools for comparing different fits, computing standard PCA, and storing results computed with foreign functions as an \class{spca} object. \pkg{spca} provides a user-friendly interface with transparent options and goodness-of-fit metrics, facilitating the production and comparison of various solutions.

The next section provides a concise introduction to the theory of LS-SPCA. Section~\ref{sec:spcapack} details the package, initially offering computational insights and subsequently presenting a comprehensive SPCA analysis through examples. As this article also functions as an extended package vignette, all main functions, their options, and helper functions are demonstrated. Tables in the illustration section are displayed as they appear in an interactive \proglang{R} session. At the end of the section, results of different LS-SPCA fits are briefly compared. In Section~\ref{compSpca}, a brief comparison with a conventional SPCA method is provided. 
The computational performance results , compared with those of a fast conventional SPCA package, are presented in Section~\ref{compPerf}. Finally, Section~\ref{conc} offers concluding remarks.

\section{LS-SPCA Methodology}\label{sec:spcamet}
The content of this section is derived from \citet{mer15} and \citet{mer19}, where comprehensive proofs and additional details are available.

We begin by defining the notation. Let $X$ represent the $n \times p$ data matrix, which comprises $n$ observations on $p$ variables, all centered to have a mean of zero. Let $S \propto X\trasp X$ denote either the covariance or correlation matrix. For the purposes of this discussion, we refer to it as the covariance matrix, as the methods and software handle both identically.

The eigenvectors of $S$, denoted as $v_1, \ldots, v_p$ (where $\vert\vert v_j \vert\vert = 1$ for $j = 1, \ldots, p$), serve as the principal component (PC) loadings and are arranged in order of decreasing eigenvalues, $\lambda_1 \geq \lambda_2 \geq \cdots \geq \lambda_p$. This ordering convention applies to all eigenvectors, including generalized eigenvectors.

The sparse loading $p$-vectors, denoted by $a_j$ for $j = 1,\ldots,r$ where $r \leq p$, have unit norm such that $a\traspj a_j = 1$. Each vector $a_j$ contains only $c_j \leq p$ nonzero elements, with $c_j$ representing the \emph{cardinality} of $a_j$. The sparse principal component scores, or sPCs, are given by $t_j = Xa_j$.

A dot above a symbol signifies the operation of retaining the nonzero elements of a vector or the columns of a matrix corresponding to a set of indices $ind_j$. In \proglang{R} syntax, $\daj = a_j[ind_j]$ is the vector that includes only the $c_j$ nonzero loadings in $a_j$, ensuring that $t_j = \dXj\da_j$.
\subsection{LS-SPCA Objective}
The LS-SPCA objective is derived by applying sparsity constraints to the conventional PCA least-squares data approximation objective, thereby maximizing the variance explained (VEXP) by each PC. The objective is recursively defined as follows:
\begin{equation}\label{eq:lsscpca_obj}
\min_{a_j} \|X - t_j b_j\trasp\|^2,
\quad \text{subject to}\quad
\operatorname{card}(a_j) < p,
\quad a\traspj S a_i = 0,\ i < j;\ j = 1, \ldots, r,
\end{equation}
the vectors $b_j$ are regression coefficients that are of no interest 

In accordance with standard model selection procedures, the problem is divided into two distinct steps:
\begin{enumerate}
\item Define the method for computing the loadings for a predetermined support ($ind_j$).
\item Identify which variable selection algorithms produce effective support.
\end{enumerate}
\subsection{Computation of the LS-SPCA Solutions}  
The LS-SPCA method comprises three variants, each characterized by distinct computational complexities and levels of optimality concerning the objective in Equation~\ref{eq:lsscpca_obj}.  

These variants differ in their approach to computing sPC loadings on a fixed support. The uncorrelated LS-SPCA (uSPCA) variant achieves the optimal solution by imposing uncorrelatedness constraints. In contrast, correlated LS-SPCA (cSPCA) substitutes these constraints with orthogonal residuals. Finally, projection LS-SPCA (pSPCA) replaces the generalized eigenproblem with a projection problem.

\paragraph{uSPCA.}  
The optimal sparse loadings, $\daj$, for objective~\ref{eq:lsscpca_obj} are identified as the generalized eigenvectors associated with the largest eigenvalue, $\lambda_j$. These eigenvectors solve the following equation:
\begin{equation}\label{eq:uspca_sol}
C_j \dSjT \dSj \daj = \lambda_j \dXjT \dXj \daj,
\end{equation}
where $\dSj = S[, ind_j]$, and the matrices $C_j$ impose orthogonality constraints. Let $\ddot{S_j} =\dXjT\dXj$ represent the covariance matrix of the $j$-th subset, and $R_j = \dA\traspdr{j -1} \dSj$. Consequently, $C_j$ is defined as:
\[
C_j = I_{c_j} - \ddot{S}\matinvj{j} R\trasps{j}
\big(R_j\ddot{S}_j R\trasps{j}\big)^{-1} R_j,
\]
with the initial condition ($C_1 = I_{c_1}$).

The orthogonality constraints require that the cardinality of the loadings is at least equal to the order $j$ of the sPC, that is, $c_j \geq j$. This requirement ensures that the matrices $R_j\ddot{S}_j R\trasps{j}$ remain nonsingular. Additionally, the computational complexity of the solutions increases, and the computation may become unstable as the number of computed sPCs increases.

\paragraph{cSPCA.} 
Simpler solutions can be derived from residual matrices that are orthogonal to the previously computed sPCs. Define
\begin{equation}\label{eq:defX_def}
Q_j = \left(I_p - \frac{t_{j-1} t\trasps{j-1}}{t\trasps{j-1} t_{j-1}}\right) Q_{j-1},
\quad j = 2, 3, \ldots, r,
\end{equation}
where $Q_1 = X$. In the SPCA literature, the residual matrices $Q_j$ are referred to as \emph{deflated} matrices. Although no orthogonality constraints are applied, the employment of deflated matrices, as a direct result of the Courant--Fischer min--max theorem, ensures low correlation with the preceding sPCs. The cSPCA solution loadings are the generalized eigenvectors associated with the largest eigenvalue, determined by solving the following equation:
\begin{equation}\label{eq:cspca_sol}
\dXjT Q_j\QjT\dXj \daj = \mu_j \dXjT\dXj \daj.
\end{equation}
The first cSPCA solution coincides with the first uSPCA solution. However, subsequent solutions generally differ from the corresponding uSPCA solutions and typically exhibit only mild mutual correlation, which tends to increase for higher-order sPCs. Their cardinality is not restricted by the uncorrelatedness requirements and is often lower than their order.
\paragraph{pSPCA.} The pSPCA method simplifies the computation by determining sparse loadings through the approximation of the first PC of the residual matrices $Q_j$ by sPCs. Let $z_j$ denote the $j$th residual PC. The objective function for pSPCA is expressed as follows:
\begin{equation}\label{eq:pscpca_obj}
\max_{a_j} \text{cor}(t_j, z_j)^2,
\quad \text{subject to}\:\text{card}(a_j) < p.
\end{equation}
The optimal solution for this objective involves linear regression, projecting the residual PCs onto a subset of variables. Consequently, in pSPCA, the sparse loadings are the coefficients from the regression of the residual PCs onto the variable subsets $\dXj$, standardized to unit norm. Specifically,
\begin{equation}
\daj \propto (\dXjT\dXj)^{-1} \dXjT z_j.
\end{equation}

Generally, the loadings obtained from pSPCA are similar to those from cSPCA, although with slightly higher mutual correlation, which remains small. This method is computationally more straightforward and offers an efficient approach to variable selection in LS-SPCA.

In the uSPCA solution, as given by Equation~\ref{eq:uspca_sol}, the eigenvalue represents the variance explained (VEXP) by the corresponding sPC. Conversely, in the cSPCA solution, as indicated by Equation~\ref{eq:cspca_sol}, the eigenvalue is typically a close approximation of the VEXP. For both cSPCA and pSPCA, the exact VEXP is calculated by the additional variance explained by each sPC, akin to linear regression, owing to the potential correlation among sPCs.
\subsection{Variable selection} 
Variable selection is a computationally demanding task classified as NP-hard, necessitating the use of greedy algorithms. Several methodological choices are available for this purpose. For uSPCA and cSPCA, we implemented intensive forward selection algorithms based on the approximate VEXP of cSPCA. This algorithm selects variables by computing the approximate cumulative variance explained (CVEXP) for each candidate variable. This approach is computationally intensive because of the requirement for repeated generalized eigendecompositions. Additionally, we implemented efficient regression-based forward, backward, and forward-stepwise variable selection algorithms derived from pSPCA. These algorithms consistently employ the computationally inexpensive $R^2$ to select or eliminate variables, with the option of using either $R^2$ or CVEXP as the stopping rule. In the latter scenario, the current CVEXP is computed at each search step only to evaluate the stopping rule, offering a computationally less demanding alternative to intensive selection. In all versions of variable selection, for efficiency, the stopping criterion CVEXP is computed using the approximate VEXP produced by cSPCA. 
\section{The spca package}\label{sec:spcapack}
The \pkg{spca} package was developed to compute and compare various sparse PCA solutions within a unified interface. It offers two computational backends: one for tall matrices ($n > p$) and another for fat matrices ($n < p$). The appropriate backend is selected automatically upon the provision of a data matrix, although users have the option to specify a particular backend. The function argument names are designed to be intuitive, with additional guidance available on the help pages and further elaboration in the \code{Details} sections. For character arguments, selecting an option requires only the first letter; if a vector is provided, as in the default settings, only the first element is used.

The primary fitting function, \fct{spca}, accept either a data matrix or a covariance matrix as input. A square symmetric matrix is interpreted as a covariance matrix; otherwise, the input is considered to be a data matrix. The fat-matrix backend requires the original data matrix. When a data matrix is provided, the package also calculates the PCs scores; otherwise, the scores are not computed.

The fitted model is returned as an object of class \class{spca}. This object retains the loading and percentage contribution matrices, the indices of nonzero loadings for each sPC, the cardinality of each sPC, as well as the VEXP and CVEXP of the sPCs, and their proportion relative to the PCs. Additionally, it contains the correlation matrix of the sPCs and the squared correlations \code{r2} between each sPC and its corresponding PC. If scores are computed, they are also stored within the object. The \class{spca} object is accompanied by its own documentation page. The package also provides auxiliary functions and methods for examining, printing, and plotting the fitted objects. Methods that generate formatted tables can return the underlying tables. Plotting functions offer a limited set of essential graphical parameters and can return \pkg{ggplot2} objects if requested, allowing users to further refine the plots using standard \pkg{ggplot2} commands.
\subsection{Computational details} 
The \pkg{spca} package is designed to retain its efficiency when computing sPCs from large matrices. Linear algebra and matrix computations are executed by the \proglang{C++} backend, which employs referenced arrays to circumvent the deep copies that are typically generated by \proglang{R} functions. The exported \proglang{C++} functions are invoked using \proglang{R} wrappers.

The primary function, \fct{spca()}, computes LS-SPCA solutions across all types. Distinct backends are utilized for handling \emph{fat} ($n < p$) and \emph{tall} matrices. The backend for tall matrices operates on the covariance matrix $S$, whereas the backend for fat matrices functions on the row space of the data matrix. This is achieved through a \emph{reverse-SVD} method to perform eigendecompositions and employs specialized variable selection algorithms to derive LS-SPCA solutions.

The primary computational strategy involves calculating the potentially intensive $O(np^2)$ covariance matrix $S = X\trasp X$ and the $O(p^3)$ product $SS$ a single time at the outset. Subsequent deflations are executed using efficient rank-1 updates, rather than recalculating the matrices as products of the deflated data matrix. This approach can lead to significant computational savings when $p$ is large, effectively reducing the cost from approximately $O(p^3)$ to $O(p^2)$.

Beyond the initial eigendecomposition, only the leading eigen-pair of a matrix is required. This can be optionally computed using the power method by setting \code{pm_loading = TRUE} for loading computation and \code{pm_varsel = TRUE} for variable selection, which is particularly effective for intensive and backward algorithms. For a $p \times p$ matrix, the power method reduces the computational cost from approximately $O(p^3)$ to approximately $O(Kp^2)$, where $K$ denotes the number of iterations necessary for convergence. Convergence is managed by the \code{maxiter_pm...} and \code{eps_pm...} parameters for each computation set.

The regression-based variable selection algorithm is simplified by recognizing that the regressed variable is a singular vector of the regressor matrix. Additionally, rank-1 updates are employed to compute the current values of the selected stopping criterion. Similar considerations are applicable to the fat backend.

\subsection{Application and illustration}
To illustrate a standard SPCA workflow, we use the classic Holzinger-Swineford dataset, available from the \proglang{R} package \pkg{psychTools} \citep{psychTools}. To keep the results compact, we use a subset of $n = 145$ observations and $p = 12$ variables, as done in other analyses \citep[for example][]{fer}. This dataset, with variables centered and scaled to unit variance, is included in the package as \code{"holzinger"}, together with a \code{factor}, \code{"holzinger_scales"}, indicating the scale to which each of the 12 variables belongs.

In evaluating and comparing LS-SPCA solutions, we prioritize percentage contributions, which are loadings scaled to a unit $L_1$ norm (sum of absolute values), rather than the conventional loadings scaled to a unit $L_2$ norm (sum of squared values). This preference arises from the enhanced interpretability of the contributions. The \pkg{spca} methods print and plot these contributions by default. Loadings are obtained by configuring \code{contributions = FALSE}.

Data are loaded as follows
\begin{knitrout}
\definecolor{shadecolor}{rgb}{0.969, 0.969, 0.969}\color{fgcolor}\begin{kframe}
\begin{alltt}
\hldef{R> }\hlkwd{data}\hldef{(}\hlsng{"holzinger"}\hldef{)}
\hldef{R> }\hlkwd{dim}\hldef{(holzinger)}
\end{alltt}
\begin{verbatim}
[1] 145  12
\end{verbatim}
\begin{alltt}
\hldef{R> }\hlkwd{data}\hldef{(}\hlsng{"holzinger_scales"}\hldef{)}
\end{alltt}
\end{kframe}
\end{knitrout}

\subsubsection*{pca()}
The \fct{pca} function executes eigendecomposition on the covariance matrix, with the results stored in an \class{spca} object. This function wraps a specialized \proglang{C++} routine that computes the truncated inverse singular value decomposition (SVD) applicable to fat matrices. Although PCA is not strictly necessary, it serves as a valuable tool for analyzing the solutions to be approximated and determining the optimal number of components to retain. The \fct{pca} function accepts the following arguments:
\begin{Code}
pca(M, n_comps = NULL, center_data = FALSE, scale_data = FALSE,
    fat_matrix = NULL, screeplot = FALSE, qq_plot = TRUE, nrow_data = NULL,
    neigen_toplot = NULL, pm = FALSE, eps_pm = 1e-05, maxiter_pm = 100)
\end{Code}
\code{M} can be either a data or a covariance matrix.
If a covariance matrix is passed, the PC scores are not computed.
The default \code{n_comps = NULL} implies that all possible components are computed.

If \code{pm = TRUE}, only \code{n_comps < p} PCs are computed using the power method, which is controlled by the convergence tolerance \code{eps_pm} and the maximum iterations permitted \code{maxiter_pm}. The function returns an \class{spca} object.

The \fct{pca} function can generate a traditional screeplot and a Wachter qq-plot \citep{wac}. The latter diagnostic plot, as explained by \cite{win}, compares the observed eigenvalues with the quantiles of the Marchenko--Pastur distribution. This distribution represents the limiting distribution of eigenvalues of random covariance matrices of independent variables with a common variance. Therefore, the Wachter qq-plot is applicable exclusively to the eigenvalues of the covariance matrices of variables with the same variance, that can be indicated by the argument \code{common_var}. However, it can be modified for use with generic correlation matrices by setting \code{common_var = 1}. Because the Marchenko--Pastur distribution is dependent on the data aspect ratio, it is necessary to specify \code{nrow_data} when \code{M} is a covariance matrix. The functions \fct{screeplot} and \fct{wachter\_qqplot} are provided for customizing these plots.

\begin{figure}[H]
\centering
\includegraphics[width = 0.45\textwidth]{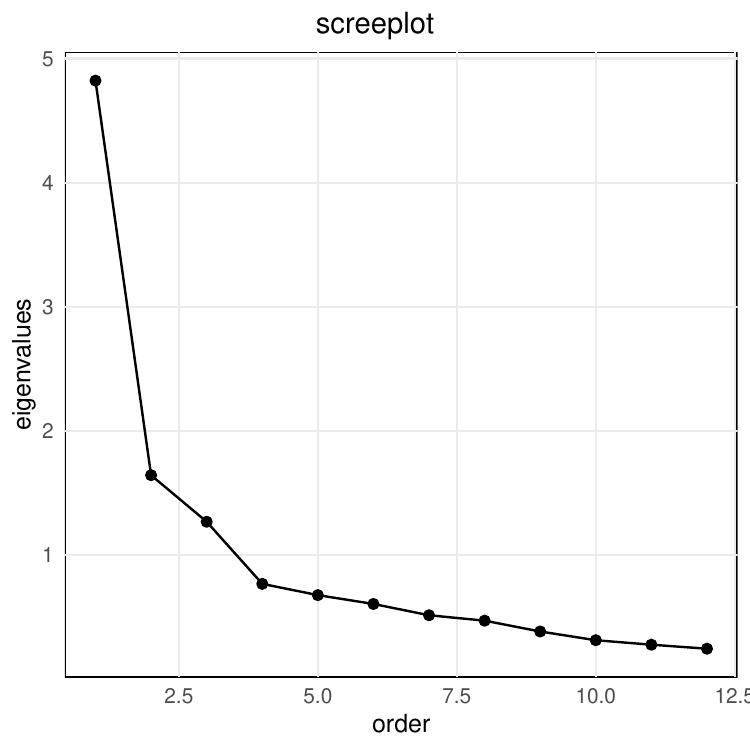}
\includegraphics[width = 0.45\textwidth]{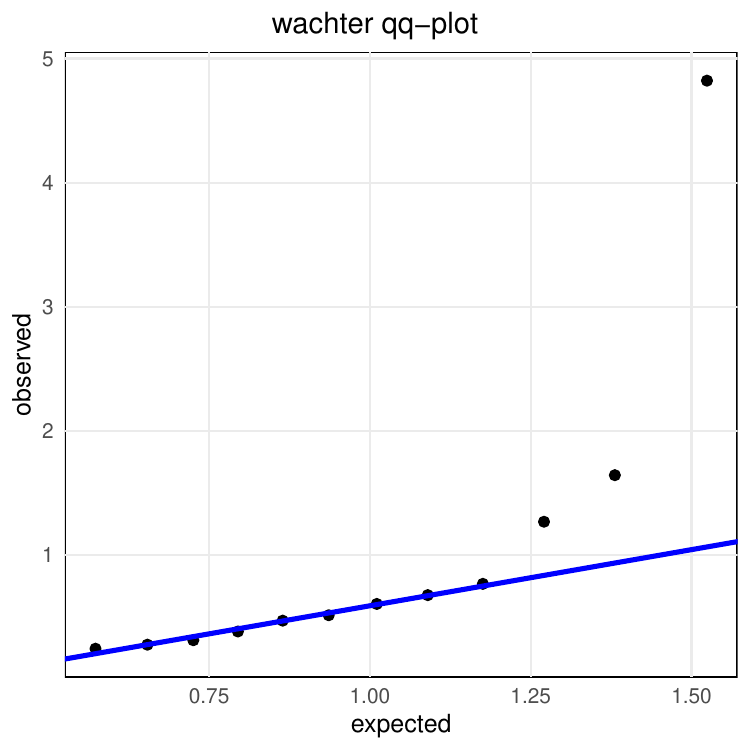}
\caption{Screeplot and qq-plot with a fitted line.}
\label{fig:pcaplots}
\end{figure}

Figure~\ref{fig:pcaplots} presents the scree and qq plots. The latter is generated independently, with a line fitted to the smallest $(n - 3)$ eigenvalues.
\begin{knitrout}
\definecolor{shadecolor}{rgb}{0.969, 0.969, 0.969}\color{fgcolor}\begin{kframe}
\begin{alltt}
\hldef{R> }\hldef{ho_pca} \hlkwb{=} \hlkwd{pca}\hldef{(holzinger,} \hlkwc{screeplot} \hldef{=} \hlnum{TRUE}\hldef{,} \hlkwc{qq_plot} \hldef{=} \hlnum{FALSE}\hldef{)}
\end{alltt}
\end{kframe}
\end{knitrout}

\begin{knitrout}
\definecolor{shadecolor}{rgb}{0.969, 0.969, 0.969}\color{fgcolor}\begin{kframe}
\begin{alltt}
\hldef{R> }\hlkwd{wachter_qqplot}\hldef{(ho_pca}\hlopt{$}\hldef{eigenvalues,} \hlkwc{p} \hldef{=} \hlkwd{ncol}\hldef{(holzinger),}
\hldef{+  }               \hlkwc{n} \hldef{=} \hlkwd{nrow}\hldef{(holzinger),} \hlkwc{n_fitline} \hldef{=} \hlopt{-}\hlnum{3}\hldef{)}
\end{alltt}
\end{kframe}
\end{knitrout}

The screeplot indicates the retention of four components, whereas the qq-plot suggests three. We chose to retain four components, recognizing that the fourth principal component may explain mainly noise.
\subsubsection*{spca()}
The main function of the package, \fct{spca}, integrates the \proglang{C++} engines to compute LS-SPCA solutions and stores the results in an \class{spca} object. This function accepts multiple parameters, as shown below,
\begin{Code}
spca(M, alpha = 0.95, n_comps = NULL, ncomp_by_cvexp = NULL,
    method = c("cspca", "uspca", "pspca"),
    var_selection = c("fwd", "bkw", "step"),
    objective = c("cvexp", "r2"), intensive = FALSE,
    fat_matrix = NULL, fixed_index_list = list(),
    center_data = FALSE, scale_data = FALSE,
    pm_loading = FALSE, eps_pm_loading = 1e-05,
    maxiter_pm_loading = 100, pm_varsel = FALSE,
    eps_pm_varsel = 1e-05, maxiter_pm_varsel = 200)
\end{Code}
Note that arguments requiring a character input can accept either a single string or a vector of strings with only the first character of the first element of the input  being  considered.

Next, we fit an LS-SPCA model with four components (\code{n_comps = 4}) using the data matrix \code{M} with default values.
\begin{knitrout}
\definecolor{shadecolor}{rgb}{0.969, 0.969, 0.969}\color{fgcolor}\begin{kframe}
\begin{alltt}
\hldef{R> }  \hldef{ho_spcadef} \hlkwb{=} \hlkwd{spca}\hldef{(}\hlkwc{M} \hldef{= holzinger,}
\hldef{+  }                  \hlkwc{alpha} \hldef{=} \hlnum{0.95}\hldef{,}          \hlcom{# 95% CVEXP}
\hldef{+  }                  \hlkwc{n_comps} \hldef{=} \hlnum{4}\hldef{,}           \hlcom{# 4 components}
\hldef{+  }                  \hlkwc{method} \hldef{=} \hlsng{"c"}\hldef{,}          \hlcom{# compute cSPCA}
\hldef{+  }                  \hlkwc{var_selection} \hldef{=} \hlsng{"f"}\hldef{,}   \hlcom{# forward selection}
\hldef{+  }                  \hlkwc{objective} \hldef{=} \hlsng{"cvexp"}\hldef{,}   \hlcom{# stop by cvexp }
\hldef{+  }                  \hlkwc{intensive} \hldef{=} \hlnum{FALSE}      \hlcom{# select by r-squared}
\hldef{+  }                \hldef{)}
\end{alltt}
\end{kframe}
\end{knitrout}
Certain parameters are explicitly defined in the displayed call for clarity, the identical outcome could be achieved with the command \code{ho_spcadef = spca(M = holzinger, n_comps = 4)}.

\subsubsection*{print()}
\fct{print} is the default method and takes the following arguments:
\begin{Code}
print.spca = function (x, cols = NULL, only.nonzero = TRUE,
                contributions = TRUE, digits = 3, thresh = 0.001,
                return_table = FALSE, components = NULL, ...
                )
\end{Code}
\fct{print} is designed to display the default percentage contributions. With the default argument \code{only_nonzero = TRUE}, variables that do not load on any sPC are omitted from the output table. The \code{cols} argument specifies which columns are printed; by default, all columns are included. If an integer is specified, the first  \code{cols} columns are printed.
\begin{knitrout}
\definecolor{shadecolor}{rgb}{0.969, 0.969, 0.969}\color{fgcolor}\begin{kframe}
\begin{alltt}
\hldef{R> }\hldef{ho_spcadef}
\end{alltt}
\end{kframe}
\end{knitrout}
\begin{kframe}

{\ttfamily\noindent\itshape\color{messagecolor}{Contributions (\%)}}\end{kframe}\begin{table}[H]
\centering
\begin{minipage}{0.95\linewidth}
\begin{verbatim}
            sPC1   sPC2   sPC3   sPC4
visual     11.9%                43.9%
cubes                    31.4% -21.3%
flags      14.2%         23.0%       
paragraph        -21.6%              
sentence   19.6%        -29.7%       
wordm            -22.3%              
addition   12.2%  27.5% -15.9% -10.6%
counting          28.6%              
straight   12.3%                     
deduct     13.7%               -24.3%
series     16.1%                     
           -----  -----  -----  -----
Cvexp      38.6%  51.5%  61.4%  67.5%
 

\end{verbatim}
\end{minipage}
\caption{Nonzero percentage contributions of the spca fit.}\label{tab:print}
\end{table}

\subsubsection*{summary()}
The \fct{summary} method produces metrics for an \pkg{spca} object, aiding in its assessment relative to the PCs. When the parameter \code{cor_with_pc = TRUE} is enabled, the output table includes the correlations between the sPCs and their corresponding PCs.

The metrics produced are as follows:
\begin{itemize}
\item \code{Vexp} The percentage variance explained.
\item \code{Cvexp} The percentage cumulative variance explained.
\item \code{Rvexp} The variance explained relative to the corresponding PC.
\item \code{Rcvexp} The cumulative variance explained relative to the
corresponding PCs.
\item \code{Card}  The number of nonzero loadings.
\item \code{r} The correlations between sPCs and corresponding PC.
\end{itemize}

\begin{knitrout}
\definecolor{shadecolor}{rgb}{0.969, 0.969, 0.969}\color{fgcolor}\begin{kframe}
\begin{alltt}
\hldef{R> }\hlkwd{summary}\hldef{(ho_spcadef,} \hlkwc{cor_with_pc} \hldef{=} \hlnum{TRUE}\hldef{)}
\end{alltt}
\end{kframe}
\end{knitrout}

\begin{table}[H]
\centering
\begin{minipage}{0.95\linewidth}
\begin{verbatim}
        sPC1  sPC2  sPC3  sPC4
Vexp   38.6% 12.9%  9.9%  6.1%
Cvexp  38.6% 51.5% 61.4% 67.5%
Rvexp  96.0% 94.5% 93.5% 95.3%
Rcvexp 96.0% 95.6% 95.3% 95.3%
Card       7     4     4     4
r      0.978 0.946 0.925 0.762

\end{verbatim}
\end{minipage}
\caption{Summaries of the spca fit.}\label{tab:sum_spca}
\end{table}

The cardinality of the loadings is significantly smaller than the number of variables (12), with all sPCs accounting for over 95\% of the VEXP of the corresponding PCs. This observation holds true for CVEXP, as expected. The first three sPCs display a strong correlation with their respective PCs. Conversely, the correlation between the fourth sPC and its corresponding PC is weaker, likely due to a reduced signal, as illustrated by the qq-plot in Figure~\ref{fig:pcaplots}. Table~\ref{tab:spc_cor} presents the correlation matrix of the sPCs.
\begin{table}[H]
\centering
\begin{minipage}{0.95\linewidth}
\begin{verbatim}
      sPC1  sPC2  sPC3  sPC4
sPC1  1.00 -0.01 -0.03 -0.02
sPC2 -0.01  1.00 -0.08 -0.10
sPC3 -0.03 -0.08  1.00 -0.06
sPC4 -0.02 -0.10 -0.06  1.00

\end{verbatim}
\end{minipage}
\caption{Mutual correlation among the sPCs.}\label{tab:spc_cor}
\end{table}

While the sPCs)were derived using cSPCA, , which allows for correlated sPCs, the resulting correlations among the sPCs are minimal. Notably, the fourth sPC exhibits a higher correlation, confirming that it conveys little signal.

\subsubsection*{plot()}
\begin{figure}[H]
\centering
\includegraphics[width = 0.6\textwidth]{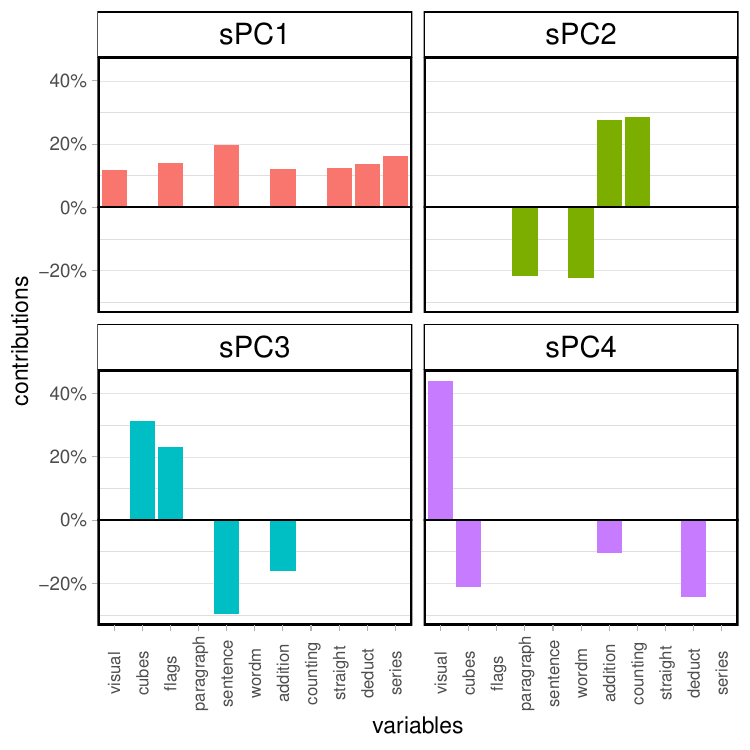}
\caption{Bar plots of the contributions of each sPC.}
\label{fig:barplot}
\end{figure}

\begin{figure}[H]
\centering
\includegraphics[width = 0.4\textwidth]{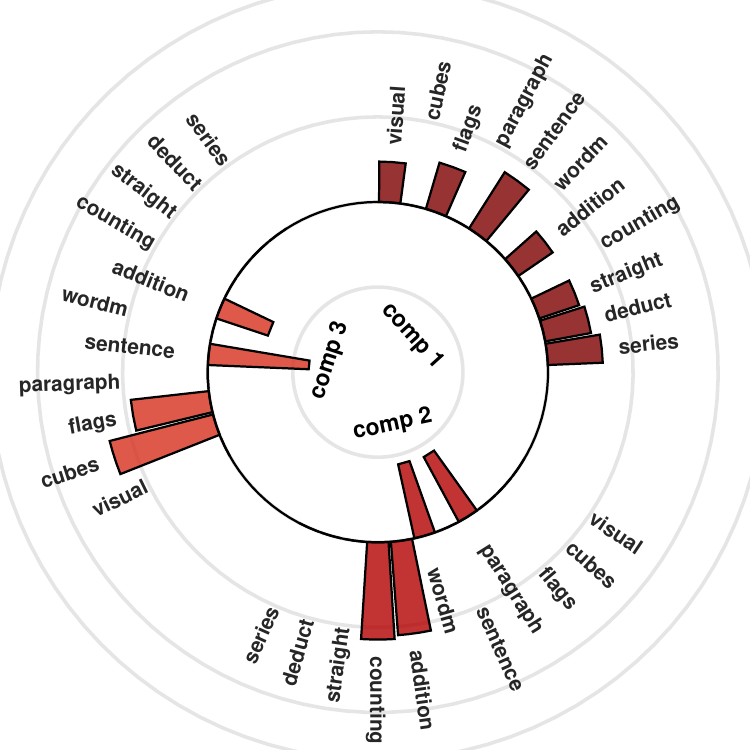}
\caption{Circular bar plots of the contributions of each sPC.}
\label{fig:circplot}
\end{figure}

The plots produced by the method \fct{plot} can be customized using various arguments. The arguments grouped under \code{controls} are \pkg{ggplot2} parameters and are included for convenience. The \pkg{ggplot2} plots can be changed manually after they are saved by setting the argument \code{return_plot} to \code{TRUE}.
\begin{Code}
plot.spca = function (
 x,
 nplot = NULL,
 plot_type = c("bars", "circular", "heatmap"),
 contributions = TRUE,
 only_nonzero = TRUE,
 pc_loadings = NULL,
 variable_groups = NULL,
 plot_title = NULL,
 return_plot = FALSE,
 produce_plot = TRUE,
 controls = list(
 color_scale = c("ggplot", "cbb", "printsafe", "bw"),
 variable_names = NULL,
 legend_position = c("none", "bottom", "right", "top", "left"),
 grid_type = c("horizontal", "full", "none"),
 facet_labels = NULL,
 legend_title = NULL,
 x_axis_lab = "variables",
 adjust_labels_circ = NULL,
 flip_heatmap = FALSE,
 heatmap_color_range = c("values", "unit")),
 ...)
\end{Code}
The argument names are self-explanatory.
When a set of PC loadings is provided to \code{pc_loadings}, the resulting plots facilitate a comparison with the sPCs loadings, except in the case of circular bar plots, where this feature is unavailable. If a vector or factor indicating groups of variables is supplied to \code{variable_groups}, the plots depict the groups accordingly. 
The \fct{plot} function generates three types of plots, producing a classic bar plot by default, as shown in Figure~\ref{fig:barplot}.
\begin{knitrout}
\definecolor{shadecolor}{rgb}{0.969, 0.969, 0.969}\color{fgcolor}\begin{kframe}
\begin{alltt}
\hldef{R> }\hlkwd{plot}\hldef{(ho_spcadef)}
\end{alltt}
\end{kframe}
\end{knitrout}
Figure~\ref{fig:circplot} presents a more compact circular plot, achieved by specifying \code{plot_type = "c"}, short for \code{"circular"}. The parameter \code{color_scale = "printsafe"} guarantees that the tones are distinguishable in a grayscale print.
\begin{knitrout}
\definecolor{shadecolor}{rgb}{0.969, 0.969, 0.969}\color{fgcolor}\begin{kframe}
\begin{alltt}
\hldef{R> }\hlkwd{plot}\hldef{(ho_spcadef,} \hlkwc{n_plot} \hldef{=} \hlnum{3}\hldef{,} \hlkwc{plot_type} \hldef{=} \hlsng{"c"}\hldef{,} \hlkwc{controls} \hldef{=} \hlkwd{list}\hldef{(}\hlkwc{color_scale} \hldef{=} \hlsng{"printsafe"}\hldef{))}
\end{alltt}
\end{kframe}
\end{knitrout}

The third option for visualizing contributions is to generate a heat map, as illustrated in the call displayed below. In this example, we incorporated the contributions of the PCs into the call, which enables a visual comparison of the two sets of contributions. The default color range was retained to vary within the range of values (\code{heatmap_color_range = "values"}), rather than between -1 and 1, as the plotted values typically span a smaller interval, rendering the colors indistinguishable.


\begin{figure}[H]
\centering
\includegraphics[width = 0.45\textwidth]{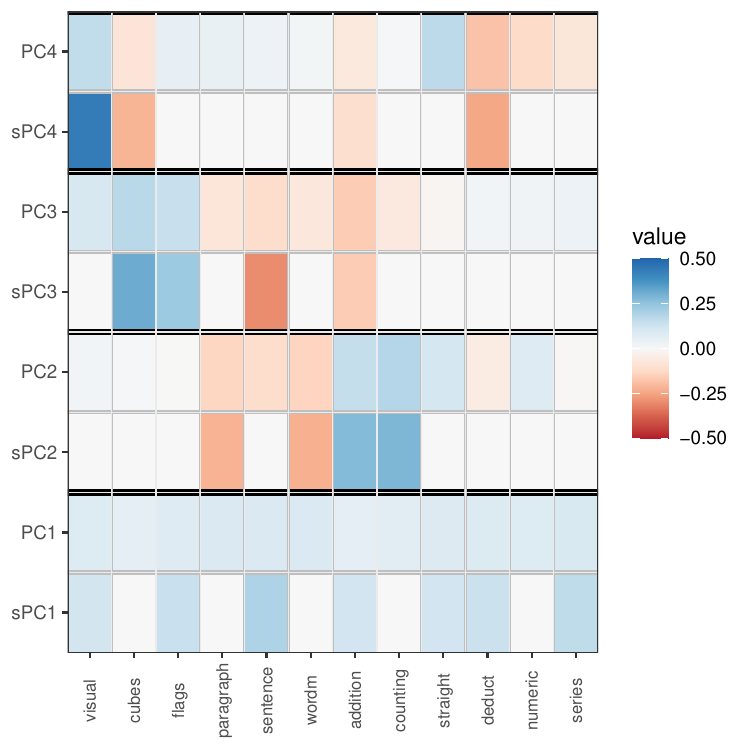}
\caption{Heat maps of the contributions of each sPC compared with the corresponding PCs contributions.}
\label{fig:heatmap}
\end{figure}

\subsubsection*{Fixed indices} 
The \code{fixed_index_list} argument within the \fct{spca} function enables the specification of variable subsets for each sPC loading set. This list must contain \code{n_comps} index vectors that collectively partition the variables in the set.

The variables in the \code{holzinger} dataset are categorized into four distinct scales, indexed by the factor \code{holzinger_scales}. A four-component \class{spca} solution can be obtained, ensuring that each sPC loads exclusively on the variables corresponding to each scale. The percentage contributions and summaries are presented in Tables~\ref{tab:fixload} and \ref{tab:fixsum}, respectively.

\begin{knitrout}
\definecolor{shadecolor}{rgb}{0.969, 0.969, 0.969}\color{fgcolor}\begin{kframe}
\begin{alltt}
\hldef{R> }\hldef{ho_spcafixed} \hlkwb{=} \hlkwd{spca}\hldef{(holzinger,} \hlkwc{alpha} \hldef{=} \hlnum{0.95}\hldef{,} \hlkwc{n_comps} \hldef{=} \hlnum{4}\hldef{,}
\hldef{+  }                 \hlkwc{fixed_index_list} \hldef{= holzinger_scales)}
\end{alltt}
\end{kframe}
\end{knitrout}

\begin{knitrout}
\definecolor{shadecolor}{rgb}{0.969, 0.969, 0.969}\color{fgcolor}\begin{kframe}
\begin{alltt}
\hldef{R> }\hldef{ho_spcafixed}
\end{alltt}
\end{kframe}
\end{knitrout}
\begin{kframe}

{\ttfamily\noindent\itshape\color{messagecolor}{Contributions (\%)}}\end{kframe}\begin{table}[H]
\centering
\begin{minipage}{0.95\linewidth}
\begin{verbatim}
            sPC1   sPC2   sPC3   sPC4
visual     41.7%                     
cubes      22.2%                     
flags      36.0%                     
paragraph         24.5%              
sentence          44.5%              
wordm             31.0%              
addition                 52.3%       
counting                 41.8%       
straight                  5.9%       
deduct                         -22.4%
numeric                        -41.5%
series                          36.1%
           -----  -----  -----  -----
Cvexp      27.2%  46.3%  61.2%  66.1%
 

\end{verbatim}
\end{minipage}
\caption{Contributions computed requiring that each sPC loads on a single scale.}\label{tab:fixload}
\end{table}

\begin{table}[H]
\centering
\begin{minipage}{0.95\linewidth}
\begin{verbatim}
         sPC1   sPC2   sPC3   sPC4
Vexp    27.2%  19.1%  14.9%   5.0%
Cvexp   27.2%  46.3%  61.2%  66.1%
Rvexp   67.6% 139.6% 141.0%  77.4%
Rcvexp  67.6%  85.9%  94.9%  93.3%
Card        3      3      3      3
r       0.761 -0.497 -0.414  0.344

\end{verbatim}
\end{minipage}
\caption{Summaries of the fits on distinct scales.}\label{tab:fixsum}
\end{table}

\subsubsection*{Compare solutions}\label{sec:comparison}
The function \code{compare_spca()} generates both numerical and visual comparative summaries for multiple \class{spca} objects.

\begin{figure}[H]
\centering
\includegraphics[width = 0.45\textwidth]{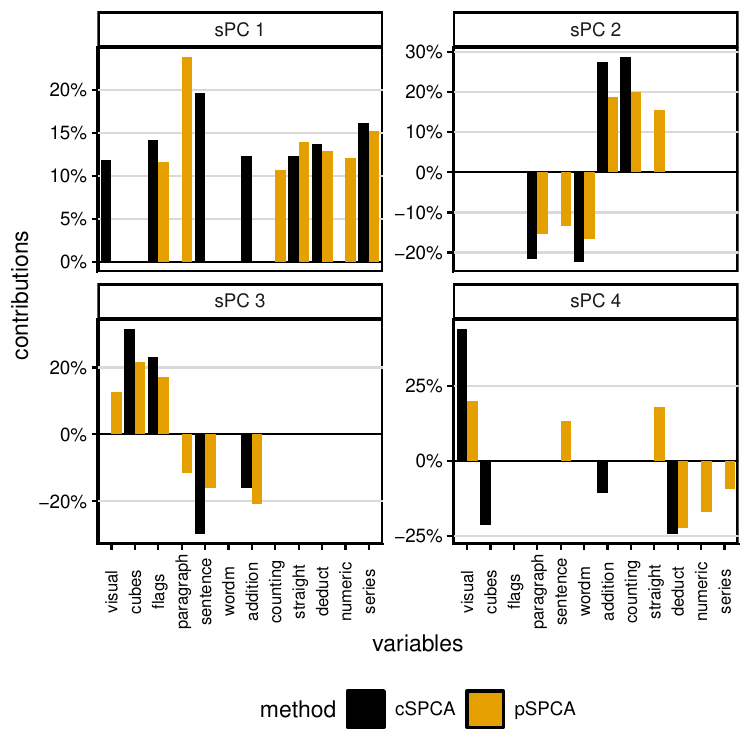}
\caption{Bar plots of the contributions of two different spca fits.}
\label{fig:cspca_vs_pspca}
\end{figure}

In this section, we compare the prior \code{cSPCA} fit with an alternative fit, where the loadings are determined using \code{pSPCA} and the variables are selected through backward elimination.
\begin{knitrout}
\definecolor{shadecolor}{rgb}{0.969, 0.969, 0.969}\color{fgcolor}\begin{kframe}
\begin{alltt}
\hldef{R> }\hldef{ho_pspca} \hlkwb{=} \hlkwd{spca}\hldef{(holzinger,} \hlkwc{n_comps} \hldef{=} \hlnum{4}\hldef{,} \hlkwc{alpha} \hldef{=} \hlnum{0.95}\hldef{,}  \hlkwc{method} \hldef{=} \hlsng{"p"}\hldef{,}
\hldef{+  }                \hlkwc{objective} \hldef{=} \hlsng{"r2"}\hldef{,} \hlkwc{var_selection} \hldef{=} \hlsng{"b"}\hldef{)}
\end{alltt}
\end{kframe}
\end{knitrout}
The method names are incorporated into the comparative bar plot in Figure~\ref{fig:cspca_vs_pspca} while maintaining the default argument \code{col_short_names = TRUE} to ensure that Table~\ref{tab:cspca_vs_pspca}, which shows the summaries for both solutions, remains compact. The parameter \code{color_scale = "cbb"} employs colors that are friendly to color-blind individuals.

\begin{knitrout}
\definecolor{shadecolor}{rgb}{0.969, 0.969, 0.969}\color{fgcolor}\begin{kframe}
\begin{alltt}
\hldef{R> }\hlkwd{compare_spca}\hldef{(}\hlkwd{list}\hldef{(ho_spcadef, ho_pspca),} \hlkwc{plot_loadings} \hldef{=} \hlnum{TRUE}\hldef{,}
\hldef{+  }             \hlkwc{color_scale} \hldef{=} \hlsng{"c"}\hldef{,}
\hldef{+  }             \hlkwc{print_loadings} \hldef{=} \hlnum{FALSE}\hldef{,}
\hldef{+  }             \hlkwc{col_short_names} \hldef{=} \hlnum{TRUE}\hldef{,}
\hldef{+  }             \hlkwc{methods_names} \hldef{=} \hlkwd{c}\hldef{(}\hlsng{"cSPCA"}\hldef{,} \hlsng{"pSPCA"}\hldef{)}
\hldef{+  }             \hldef{)}
\end{alltt}
\end{kframe}
\end{knitrout}

\includegraphics[width=\maxwidth]{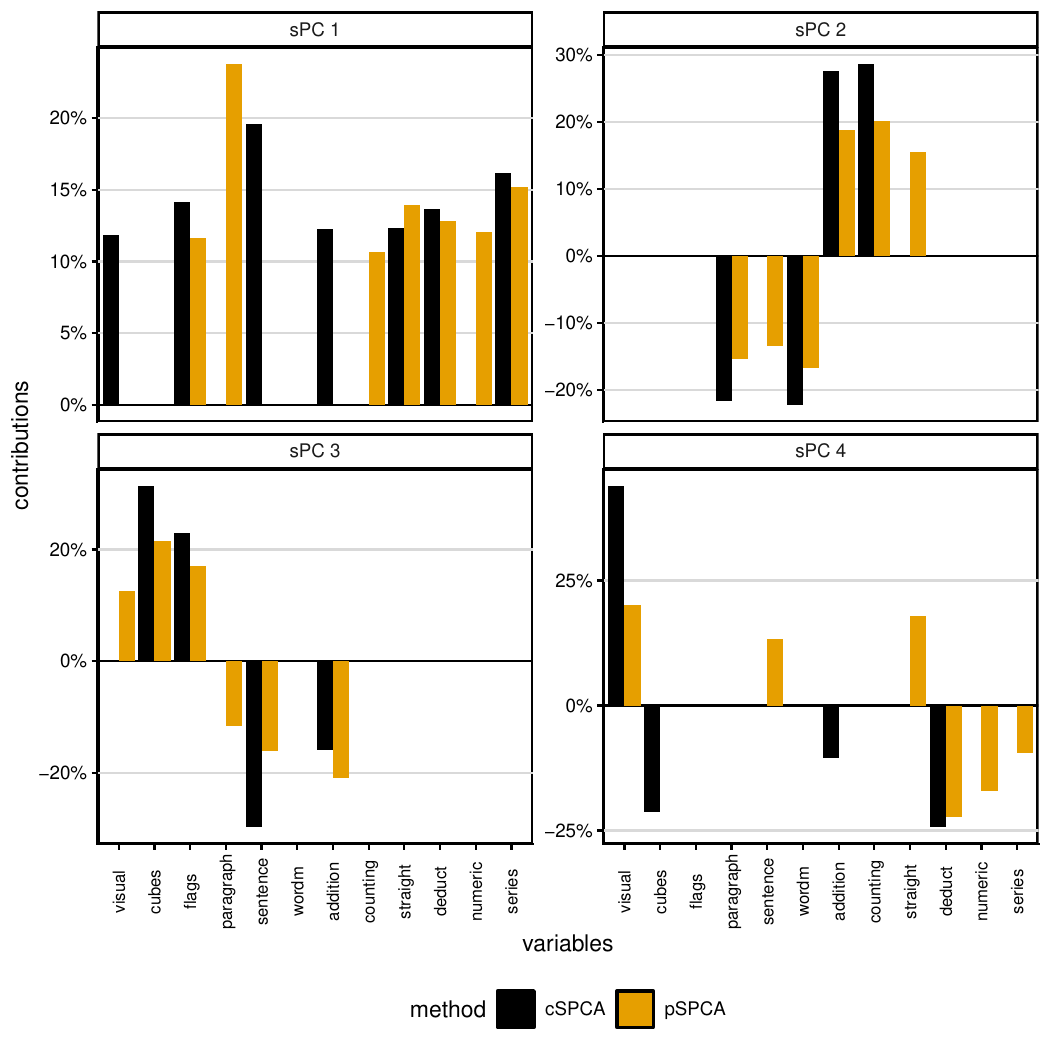} 
\begin{table}[H]
\centering
\begin{minipage}{0.95\linewidth}
\begin{verbatim}
[1] Summary statistics
       C1.M1  C1.M2  C2.M1  C2.M2  C3.M1  C3.M2  C4.M1  C4.M2 
Vexp    38.6%  38.6%  12.9%  13.5%   9.9%  10.4%   6.1%   6.4%
Cvexp   38.6%  38.6%  51.5%  52.1%  61.4%  62.5%  67.5%  68.8%
Rvexp   96.0%  96.0%  94.5%  98.5%  93.5%  97.9%  95.3%  99.8%
Rcvexp  96.0%  96.0%  95.6%  96.7%  95.3%  96.9%  95.3%  97.1%
Card        7      7      4      6      4      6      4      6
abs_r    0.98   0.98   0.95   0.99   0.92   0.98   0.76   0.94

\end{verbatim}
\end{minipage}
\caption{Comparative summaries of two different spca fits.}\label{tab:cspca_vs_pspca}
\end{table}

Apart from the first, the pSPCA loadings have larger cardinality and the  sPCs explain more variance. This is expected because VEXP cannot be lower than the $R^2$ optimized. The contributions of the first three sets of loadings are not very dissimilar, considering the different methods used, and these sPCs are also highly mutually correlated. The fourth component behaves differently, confirming that it explains mostly noise.
\subsubsection*{Variable groups}
In some cases, typically for questionnaire data, variables belong to different groups, or scales. The 12 variables in the \code{holzinger} data are grouped into four scales.

The function \fct{aggregate_by_group} produces partial sums of the loadings or contributions of an \class{spca} object by indices indicated by the vector or factor passed to the argument \code{groups}, as shown in Table~\ref{tab:aggr_by_group}.
\begin{knitrout}
\definecolor{shadecolor}{rgb}{0.969, 0.969, 0.969}\color{fgcolor}\begin{kframe}
\begin{alltt}
\hldef{R> }\hlkwd{aggregate_by_group}\hldef{(ho_spcadef,} \hlkwc{groups} \hldef{= holzinger_scales)}
\end{alltt}
\end{kframe}
\end{knitrout}
\begin{table}[H]
\centering
\begin{minipage}{0.95\linewidth}
\begin{verbatim}
         sPC1       sPC2       sPC3       sPC4
SPL 0.2603418  0.0000000  0.5439836  0.2256754
VBL 0.1957949 -0.4390701 -0.2965215  0.0000000
SPD 0.2456451  0.5609299 -0.1594949 -0.1055006
MTH 0.2982181  0.0000000  0.0000000 -0.2425778

\end{verbatim}
\end{minipage}
\caption{Contributions aggregated by scale.}\label{tab:aggr_by_group}
\end{table}

\begin{figure}[H]
\centering
\begin{knitrout}
\definecolor{shadecolor}{rgb}{0.969, 0.969, 0.969}\color{fgcolor}\begin{kframe}
\begin{alltt}
\hldef{R> }\hlkwd{plot}\hldef{(ho_spcadef,} \hlkwc{variable_groups} \hldef{= holzinger_scales,} \hlkwc{controls} \hldef{=}
\hldef{+  }       \hlkwd{list}\hldef{(} \hlkwc{legend_position} \hldef{=} \hlsng{"right"}\hldef{)}
\hldef{+  }\hldef{)}
\end{alltt}
\end{kframe}
\end{knitrout}
\includegraphics[width = 0.45\textwidth]{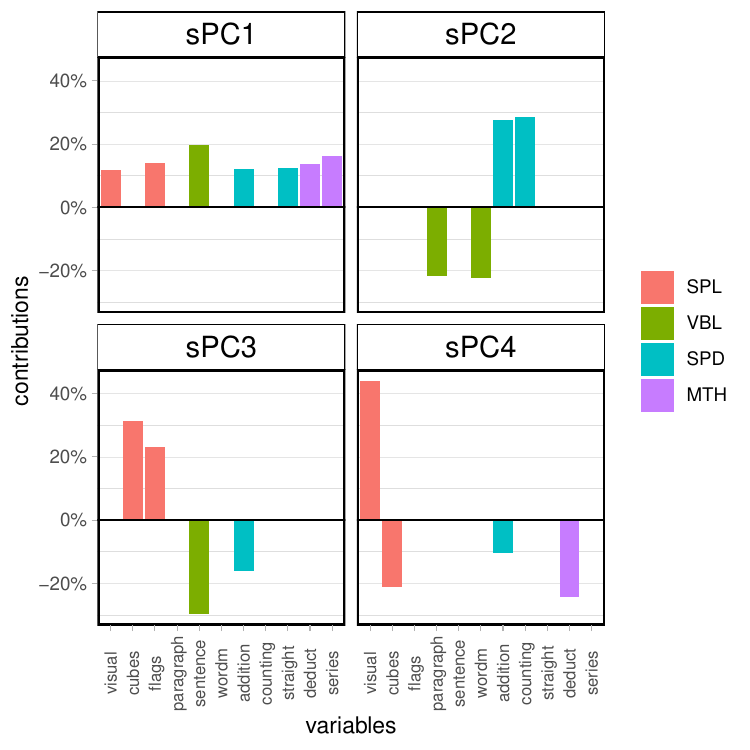}
\caption{Bar plots of the contributions filled by groups.}
\label{fig:plot_by_grp}
\end{figure}

Passing an index vector  to the argument \code{variable_groups} to the \fct{plot} method returns a plot with scales filled with different colors, as shown in Figure~\ref{fig:plot_by_grp}.
\subsubsection*{Create an \class{spca} object}
The \pkg{spca} functions can be employed on the results from other analyses by converting them into an \class{spca} object through the \fct{new_spca} function. This function requires a set of loadings and either the covariance or the data matrix.
\begin{knitrout}
\definecolor{shadecolor}{rgb}{0.969, 0.969, 0.969}\color{fgcolor}\begin{kframe}
\begin{alltt}
\hldef{R> }\hldef{A} \hlkwb{=} \hlkwd{cbind}\hldef{(ho_spcadef}\hlopt{$}\hldef{loadings[,} \hlnum{1}\hldef{], ho_pspca}\hlopt{$}\hldef{loadings[,} \hlnum{2}\hldef{])}
\hldef{R> }\hldef{ho_r} \hlkwb{=} \hlkwd{cor}\hldef{(holzinger)}
\hldef{R> }\hldef{ho_spcahyb} \hlkwb{=} \hlkwd{new_spca}\hldef{(A, ho_r,} \hlkwc{method_name} \hldef{=} \hlsng{"hybrid"}\hldef{)}
\end{alltt}
\end{kframe}
\end{knitrout}
\begin{knitrout}
\definecolor{shadecolor}{rgb}{0.969, 0.969, 0.969}\color{fgcolor}\begin{kframe}
\begin{alltt}
\hldef{R> }\hlkwd{is.spca}\hldef{(ho_spcahyb)}
\end{alltt}
\begin{verbatim}
[1] TRUE
\end{verbatim}
\end{kframe}
\end{knitrout}
\section{Comparison of LS-SPCA variants}\label{sec:comparisons}
In this section, we present a concise comparison of the LS-SPCA results obtained using different computational and variable selection methods, as well as different values of \code{alpha}. For these comparisons, we employ \fct{spca} with its default settings, modifying one parameter at a time to isolate the effect of that specific parameter.

The analyses are performed on the Multidimensional Sexual Self-Concept Questionnaire dataset \citep[MSSCQ,][]{msscqdata}. After excluding observations with more than three zero values, the dataset consists of 16,985 responses across 100 items, which we centered to a zero mean and standardized to unit variance. This dataset is referred to as \code{mss}. 

\subsubsection*{Computation methods}
Three different solutions are obtained by setting the parameter \code{method} to \code{uspca}, \code{cspca} and \code{pspca}. The summary statistics for these three solutions are presented in Table~\ref{tab:comp_meth}.
\begin{knitrout}
\definecolor{shadecolor}{rgb}{0.969, 0.969, 0.969}\color{fgcolor}\begin{kframe}
\begin{alltt}
\hldef{R> }\hldef{met} \hlkwb{=} \hlkwd{c}\hldef{(}\hlsng{"uspca"}\hldef{,} \hlsng{"cspca"}\hldef{,} \hlsng{"pspca"}\hldef{)}
\hldef{R> }\hldef{mss_met_spca} \hlkwb{=} \hlkwd{vector}\hldef{(}\hlsng{"list"}\hldef{,} \hlnum{3}\hldef{)}
\hldef{R> }\hldef{pkram} \hlkwb{=} \hlkwd{vector}\hldef{(}\hlsng{"list"}\hldef{,} \hlnum{3}\hldef{)}
\hldef{R> }
\hldef{R> }\hlkwa{for}\hldef{(i} \hlkwa{in} \hlnum{1}\hlopt{:}\hlnum{3}\hldef{)\{}
\hldef{+  }  \hldef{pkram[[i]]} \hlkwb{=} \hldef{peakRAM}\hlopt{::}\hlkwd{peakRAM}\hldef{(\{}
\hldef{+  }    \hldef{mss_met_spca[[i]]} \hlkwb{=} \hlkwd{spca}\hldef{(mss,} \hlkwc{n_comps} \hldef{=} \hlnum{4}\hldef{,} \hlkwc{method} \hldef{= met[i])}
\hldef{+  }  \hldef{\})[}\hlnum{1}\hldef{,} \hlopt{-}\hlnum{1}\hldef{]}
\hldef{+  }\hldef{\}}
\hldef{R> }
\hldef{R> }\hldef{mss_met_table} \hlkwb{=} \hlkwd{make_comparative_table}\hldef{(}\hlkwc{L} \hldef{= mss_met_spca,} \hlkwc{ind} \hldef{=} \hlnum{1}\hlopt{:}\hlnum{3}\hldef{,}
\hldef{+  }                                       \hlkwc{pRAM} \hldef{=} \hlkwa{NULL}\hldef{,} \hlkwc{par_name} \hldef{=} \hlsng{"method"}\hldef{,}
\hldef{+  }                                       \hlkwc{par_values} \hldef{= met)}
\end{alltt}
\end{kframe}
\end{knitrout}
\begin{table}[H]
\centering
\begin{minipage}{0.95\linewidth}
\begin{verbatim}
  method CVEXP             Card             Cor.with.PCs max|cor|
1  uspca 95.8% [13, 20, 21, 30] [0.98, 0.98, 0.98, 0.94]    0.000
2  cspca 95.8% [13, 20, 21, 30] [0.98, 0.98, 0.97, 0.94]    0.019
3  pspca 95.8% [13, 20, 21, 31] [0.98, 0.98, 0.98, 0.95]    0.020

\end{verbatim}
\end{minipage}
\caption{Comparative summaries for different computation methods}\label{tab:comp_meth}
\end{table}

In this example there are no noticeable differences in the three solutions; however, as required, the uSPCA sPCs are perfectly uncorrelated. 

\subsubsection*{Variable selection methods}
In this analysis, two parameters are varied: the search direction, \code{var_selection}, and stopping rule, \code{objective}. The searches employ partial squared correlation, \code{"r2"}, for the selection of candidate variables, except when the parameter \code{intensive} is set to \code{TRUE}. In such cases, CVEXP is used to select candidate variables. Table~\ref{tab:comp_varsel} presents the summary statistics for the seven variable selection approaches.
%
%
%
\begin{table}[H]
\centering
\begin{minipage}{0.95\linewidth}
\begin{verbatim}
     var sel CVEXP             Card             Cor.with.PCs max|cor|
1     fwd r2 95.8% [13, 20, 21, 30] [0.98, 0.98, 0.97, 0.94]    0.019
2    step r2 95.8% [13, 20, 21, 30] [0.98, 0.98, 0.97, 0.94]    0.019
3     bwd r2 95.7% [13, 20, 20, 30] [0.98, 0.98, 0.97, 0.95]    0.029
4  fwd cvexp 95.1% [12, 18, 16, 18] [0.97, 0.97, 0.96, 0.91]    0.034
5 step cvexp 95.0% [12, 18, 16, 16] [0.97, 0.97, 0.96, 0.88]    0.040
6  bwd cvexp 95.0% [13, 17, 15, 17] [0.97, 0.97, 0.95, 0.89]    0.029
7  intensive 95.0% [13, 16, 16, 14] [0.97, 0.97, 0.96, 0.85]    0.031

\end{verbatim}
\end{minipage}
\caption{Comparative summaries for different variable selection methods}\label{tab:comp_varsel}
\end{table}

The selection with the \code{r2} objective produces loadings with moderately higher cardinality, resulting in sPCs that are slightly more correlated with the PCs. The \code{intensive} search produces more parsimonious loadings with characteristics similar to those of the other solutions. Notably, under the "r2" objective, the fourth sPC demonstrates significantly greater cardinality than that obtained with \code{cvexp}.
\subsubsection*{alpha}
The parameter \code{alpha} determines the target for the minimum CVEXP, or the correlation with the residual PCs. A lower \code{alpha} value results in a greater difference between sPCs and PCs. However, this typically leads to a reduction in the cardinality of the loadings. Table~\ref{tab:comp_alpha} presents the summary statistics for the default solutions calculated with \code{alpha = c(0.90, 0.95, 0.98)}.
 
\begin{knitrout}
\definecolor{shadecolor}{rgb}{0.969, 0.969, 0.969}\color{fgcolor}\begin{kframe}
\begin{alltt}
\hldef{R> }\hldef{alpha} \hlkwb{=} \hlkwd{c}\hldef{(}\hlnum{0.90}\hldef{,} \hlnum{0.95}\hldef{,} \hlnum{0.98}\hldef{)}
\hldef{R> }
\hldef{R> }\hldef{mss_alpha_spca} \hlkwb{=} \hlkwd{vector}\hldef{(}\hlsng{"list"}\hldef{,} \hlnum{3}\hldef{)}
\hldef{R> }\hldef{pkram} \hlkwb{=} \hlkwd{vector}\hldef{(}\hlsng{"list"}\hldef{,} \hlnum{3}\hldef{)}
\hldef{R> }
\hldef{R> }\hlkwa{for}\hldef{(i} \hlkwa{in} \hlnum{1}\hlopt{:}\hlnum{3}\hldef{)\{}
\hldef{+  }  \hldef{pkram[[i]]} \hlkwb{=} \hldef{peakRAM}\hlopt{::}\hlkwd{peakRAM}\hldef{(\{}
\hldef{+  }    \hldef{mss_alpha_spca[[i]]} \hlkwb{=} \hlkwd{spca}\hldef{(mss,} \hlkwc{alpha} \hldef{= alpha[i],} \hlkwc{n_comps} \hldef{=} \hlnum{4}\hldef{)}
\hldef{+  }  \hldef{\})[}\hlnum{1}\hldef{,} \hlopt{-}\hlnum{1}\hldef{]}
\hldef{+  }\hldef{\}}
\hldef{R> }
\hldef{R> }\hldef{mss_alpha_table} \hlkwb{=} \hlkwd{make_comparative_table}\hldef{(}
\hldef{+  }  \hlkwc{L} \hldef{= mss_alpha_spca,} \hlkwc{pRAM} \hldef{=} \hlkwa{NULL}\hldef{,}
\hldef{+  }  \hlkwc{par_name} \hldef{=} \hlsng{"alpha"}\hldef{,} \hlkwc{par_values} \hldef{= alpha}
\hldef{+  }\hldef{)}
\end{alltt}
\end{kframe}
\end{knitrout}

\begin{table}[H]
\centering
\begin{minipage}{0.95\linewidth}
\begin{verbatim}
  alpha CVEXP             Card             Cor.with.PCs max|cor|
1  0.90 92.3%  [7, 11, 12, 20] [0.95, 0.95, 0.94, 0.84]    0.038
2  0.95 95.8% [13, 20, 21, 30] [0.98, 0.98, 0.97, 0.94]    0.019
3  0.98 98.3% [27, 35, 32, 49] [0.99, 0.99, 0.99, 0.99]    0.014

\end{verbatim}
\end{minipage}
\caption{Comparative summaries of solutions obtained with increasing values of alpha}\label{tab:comp_alpha}
\end{table}

The results confirm the expectations mentioned above. The sPCs computed with \code{alpha = 0.98} are parsimonious versions of the PCs.

\section{Comparison of LS-SPCA with conventional SPCA}\label{compSpca}
In this section, we present a comparison between LS-SPCA solutions obtained with our package \pkg{spca} and conventional SPCA solutions computed by the \proglang{R} package \pkg{elasticnet} version 4.1-8 \citep{zou, elasticnet}. The \pkg{elasticnet} package was chosen because it is the foundational conventional SPCA implementation and remains the most frequently cited benchmark. Broader comparisons with other conventional SPCA packages are beyond the scope of this package introduction; these methods optimize the same variance objective and mainly differ in penalties and optimization algorithms. We refer readers to the series of articles by \citeauthor{camI} and references therein for more extensive comparisons. For brevity, we refer to the solutions computed with our package as \code{ls-spca}, and to those computed with \pkg{elasticnet} as \code{en-spca}.

Since \pkg{elasticnet} \fct{spca} requires penalty tuning to control sparsity, which can be time-consuming and may lead to different cardinalities, we fixed the number of components and matched the cardinalities returned by the default \fct{spca}. We also disabled internal cardinality tuning by setting \code{sparse = "varnum"}, leaving the other arguments at their defaults. This allows the comparison to focus on the main distinction between the approaches: LS-SPCA optimizes the data approximation objective, whereas conventional SPCA optimizes the variance of the sPCs.

\subsubsection*{Tall matrices}\label{sec:lm_vs_ab_tall}
For the comparisons on tall matrices, four sPCs were computed on the MSSCQ dataset with cardinalities of $13, 20, 21,$ and $30$, as resulted from the default \fct{spca}.

Table~\ref{tab:ls_en-sum} shows comparative metrics for the \code{ls-spca} and \code{en-spca} default fits.


%

\begin{table}[H]
\centering
\begin{minipage}{0.95\linewidth}
\begin{verbatim}
[1] Summary statistics
       C1.ls  C1.el  C2.ls  C2.el  C3.ls  C3.el  C4.ls  C4.el 
Vexp    22.7%  20.0%   9.3%   9.3%   6.6%   6.6%   3.3%   5.7%
Cvexp   22.7%  20.0%  31.9%  29.3%  38.5%  35.9%  41.8%  41.6%
Rvexp   95.4%  84.3%  95.5%  96.4%  96.1%  96.0%  99.5% 171.4%
Rcvexp  95.4%  84.3%  95.4%  87.8%  95.5%  89.2%  95.8%  95.4%
Card       13     13     20     20     21     21     30     30
abs_r    0.98   0.90   0.98   0.72   0.97   0.80   0.94   0.12

\end{verbatim}
\end{minipage}
\caption{Comparative summaries for default spca() (ls-spca) and elasticnet spca() (en-spca)}\label{tab:ls_en-sum}
\end{table}

The \code{ls-spca} solution consistently achieves a CVEXP exceeding 95\% for all components, as required, and remains consistently higher than the CVEXP of \code{en-spca}. However, this difference decreases as the number of components increases. In contrast, the correlation between the en-sPCs and the corresponding PCs is significantly lower and decreases substantially for higher-order sPCs.

Another notable difference between these solutions is the mutual correlation between the sPCs, shown in Table~\ref{tab:lsen_spcCor}.
\begin{table}[H]
\centering
\begin{minipage}{0.95\linewidth}
\begin{verbatim}
     ls-sPC1 ls-sPC2 ls-sPC3 ls-sPC4 en-sPC1 en-sPC2 en-sPC3 en-sPC4
sPC1    1.00    0.00    0.01    0.00    1.00    0.35   -0.37    0.76
sPC2    0.00    1.00    0.02    0.01    0.35    1.00   -0.22    0.28
sPC3    0.01    0.02    1.00    0.00   -0.37   -0.22    1.00   -0.19
sPC4    0.00    0.01    0.00    1.00    0.76    0.28   -0.19    1.00

\end{verbatim}
\end{minipage}
\caption{Separate correlation matrices of the sPCs
        computed with default spca() (ls-spca) and 
        elasticnet spca() (en-spca).}\label{tab:lsen_spcCor}
\end{table}

While the ls-sPCs are nearly  uncorrelated, the en-sPCs show a much higher mutual correlation.

As a consequence of the objective adopted, conventional SPCA algorithms aim to select as many correlated variables as possible, whereas LS-SPCA tends to favor less correlated variables. Because variables in the same scale are typically highly mutually correlated, we expect the en-sPCs to load more on variables within the same scale. Table~\ref{tab:mss_ls_en_aggr} shows the contributions of \code{ls-spca} and \code{en-spca} aggregated by scale.

\begin{table}[H]
\centering
\begin{minipage}{0.95\linewidth}
\begin{verbatim}
    ls-sPC1 ls-sPC2 ls-sPC3 ls-sPC4 en-sPC1 en-sPC2 en-sPC3 en-sPC4
SAN  -8.3%    9.6%    4.1%                                  -29.3% 
SSE   8.6%                    2.9%   24.8%                    0.1% 
SC    6.2%    3.8%                            5.1%                 
MTA                   7.7%    3.2%                                 
CLS           9.0%            9.0%                           -5.3% 
SP           19.6%   -7.6%   -5.2%           29.1%                 
SAS   8.1%                   -2.3%            5.1%            2.6% 
SO    6.8%                    1.8%                   -0.7%    7.4% 
SPS           4.9%   16.6%   -4.1%                  -20.7%         
SMN           8.1%            8.1%                           -5.0% 
SMT          16.1%   -5.9%   -6.7%           41.4%                 
SPM           6.2%   17.1%   -4.7%                  -32.9%         
SE   16.7%                    7.8%   35.4%                    7.7% 
SS    9.6%                   12.1%   39.7%                    0.8% 
POS           6.4%           26.0%                           -8.1% 
SSS   7.0%    6.1%                           13.8%                 
FOS  -9.5%           10.9%    3.5%           -5.5%          -14.2% 
SPP   4.8%           18.0%                          -25.9%         
SD   -8.2%   10.2%                                          -19.5% 
ISC   6.1%           12.1%   -2.5%                  -19.8%         

\end{verbatim}
\end{minipage}
\caption{Contributions aggregated by scale.}\label{tab:mss_ls_en_aggr}
\end{table}

As expected, conventional SPCA solutions load on fewer scales than the LS-SPCA solutions.

\pkg{elasticnet} is likely to be less computationally efficient than \pkg{spca} because it is entirely written in \proglang{R}. Table~\ref{tab:ls_en_times} shows the computation times and RAM usage of a single run for matching cardinality solutions, providing an idea of their magnitudes.
                                        
\begin{table}[H]
\centering
\begin{minipage}{0.95\linewidth}
\begin{verbatim}
        Time_sec Tot RAM MiB peak RAMV MiB
ls-spca     0.45         0.1         14.80
en-spca    99.42         0.2         84.70
en/ls     220.93         2.0          5.72

\end{verbatim}
\end{minipage}
\caption{Computational times and memory allocation for spca and elasticnet and their ratio (last row)}\label{tab:ls_en_times}
\end{table}

In the two SPCA approaches, VEXP is measured differently: in conventional SPCA it is measured as the leading eigenvalue of the covariance matrix of the selected variables; in LS-SPCA it is measured by the squared Euclidean norm of the dimensionally reduced data approximation. Table~\ref{tab:ls_en_vexp} shows the different measures obtained from the respective fits.
\begin{table}[H]
\centering
\begin{minipage}{0.95\linewidth}
\begin{verbatim}
         sPC1 sPC2 sPC3 sPC4
en pev   0.07 0.07 0.06 0.04
ls vexp  0.20 0.09 0.07 0.06
pev/vexp 0.37 0.70 0.88 0.66

\end{verbatim}
\end{minipage}
\caption{Difference in relative variance explained as measured in LS-SPCA (ls vexp) and in conventional SPCA (en pev).}\label{tab:ls_en_vexp}
\end{table}

Clearly, the conventional SPCA \code{pev} is unrelated to LS-SPCA \code{vexp}.
\subsubsection*{Fat matrices}
For fat matrices where $(n < p)$, conventional SPCA methods can return loadings with cardinality larger than the matrix rank $(n - 1)$ because the variables are mean-centered \citep[see, for example.][]{zou, mog}. This is counterintuitive, as the linear dependence of columns implies that any linear combination of the variables can be expressed using at most $(n - 1)$ linearly independent variables \citep[\emph{e.g.}][]{mer19}.

We illustrate this point using the \code{gasoline} dataset, which contains 402 near-infrared  readings on 60 gasoline samples, available in the \proglang{R} package \pkg{pls} \citep{pls}. Table~\ref{tab:fatlsenab} presents a comparison between an \code{ls-spca} solution computed with \code{alpha = 0.999} and solutions derived from \pkg{elasticnet}'s \fct{spca} and \pkg{abess}'s \fct{abesspca}, both restricted to a cardinality of 100.
%
%
%
%
%

\begin{table}[H]
\centering
\begin{minipage}{0.95\linewidth}
\begin{verbatim}
[1] Summary statistics
       C1.ls  C1.en  C1.ab  C2.ls  C2.en  C2.ab 
Vexp    71.5%  71.6%  69.0%  16.8%  16.8%  15.6%
Cvexp   71.5%  71.6%  69.0%  88.4%  88.4%  84.7%
Rvexp  100.0% 100.0%  96.4%  99.9% 100.0%  93.0%
Rcvexp 100.0% 100.0%  96.4% 100.0% 100.0%  95.8%
Card        6    100    100      8    100    100
abs_r    1.00   1.00   0.98   1.00   1.00   0.51

\end{verbatim}
\end{minipage}
\caption{Summary statistics comparing cspca with alpha = 0.999 solutions to
two conventional SPCA solutions, one computed with elasticnetspca (Cx-en) and the other with abess abesspca (Cx.ab), both requiring cardinality 100.}\label{tab:fatlsenab}
\end{table}

Conventional SPCA methods compute components with redundant variables; in this example, the maximum meaningful cardinality is 59, yet both \code{en-spca} and \code{ab-spca} compute loadings with a cardinality of 100. Conversely, \code{ls-spca} achieves 0.999 CVEXP with cardinalities of 6 and 8.

Both \code{ls-spca} and \code{en-spca} almost perfectly recover the first two PCs, with RCVEXP equal to 100\% and correlations equal to 1. The \code{ab-spca} solution is less accurate: the second sPC reaches only 95.8\% RCVEXP and has a correlation of 0.51 with the second PC. Moreover, its two sPCs are still strongly correlated, with a correlation of 0.84, whereas the \code{ls-spca} and \code{en-spca} components are effectively uncorrelated because they are perfectly aligned with the PCs.

\section{Computational performance}\label{compPerf}
The most expensive operation in \fct{spca} is typically the first eigendecomposition. However, some variable selection algorithms are more computationally demanding than others. The complexity of these searches depends on several parameters, such as the number of variables selected. The  dominant terms, which omit lower-order terms, are listed in Table~\ref{tab:comp_compl}. These are intended only to compare the scaling of the selection strategies.

Let \(n\) be the number of observations, \(p\) the number of variables,
\(r\) the number of components in the model. Let \(c\) denote the target
cardinality, or number of variables selected.
Also, let \(I_{\mathrm{step}}\) denote the number of stepwise iterations, that is, the number of add/delete or exchange updates evaluated during the stepwise selection. Thus, \(I_{\mathrm{step}}\) is algorithm-dependent and not necessarily equal to \(c\).
\begin{table}[H]
\[
\begin{array}{ll}
\text{Tall, r2 forward:} & O(r n p c) \\
\text{Tall, r2 stepwise:} & O(r I_{\mathrm{step}} n p c) \\
\text{Tall, r2 backward:} & O(r n p^2) \\
\text{Tall, cvexp forward:} & O(r p n c^3 + r p c^4) \\
\text{Tall, intensive cvexp forward:} & O(r p n c^4 + r p c^5) \\
\text{Tall, cvexp stepwise:} & O(r I_{\mathrm{step}} p n c^3 + r I_{\mathrm{step}} p c^4) \\
\text{Tall, cvexp backward:} & O(r n p^4 + r p^5) \\
\text{Fat, r2 forward:} & O(r n p c) \\
\text{Fat, cvexp forward:} & O(r p n^2 c + r p n c^3 + r p c^4)
\end{array}
\]
\caption{Dominant terms of the computational complexity of the variable
selection algorithms implemented in spca}
\label{tab:comp_compl}
\end{table}
Notably, forward and stepwise algorithms are linear in $p$, regardless of the
objective. In contrast, backward algorithms scale as $p^2$ for the $r2$
objective and as $p^5$ for \code{cvexp}. Moreover, when \code{cvexp} is used,
the dependence on the cardinality $c$ becomes much stronger, with dominant
terms involving $c^3$, $c^4$, and, for the intensive and backward variants,
$c^5$.
\subsection{Scalability across matrix dimensions}
\fct{spca} scalability was assessed by running 100 benchmark iterations for matrices with up to 2,000 variables. For fat matrices, we vary the number of observations because the dominant computational cost is the eigendecomposition in the row space; the number of variables affects the selection step, which is linear in \(p\) under forward selection. performance was measured on a Windows PC with an Intel(R) Core(TM) i7-8750H CPU @ 2.20GHz (2.21 GHz) CPU and 8Gb of RAM.

Data were generated according to a simple algorithm that guarantees a factorial structure, as shown below.
\begin{Code}
generate_data = function(n, r, p, sigma, stand = TRUE, seed = 1){
  set.seed(seed)
  M = matrix(rnorm(n * r), n, r)            #signal
  B = matrix(rnorm(r * p, 0, 0.5), r, p)    # r x p loading matrix
  E = matrix(rnorm(n * p, 0, sigma), n, p)  # noise
  X = M 
  if(stand)
    scale(X, TRUE, TRUE)
  else
    scale(X, TRUE, FALSE)
}
\end{Code}
Table~\ref{tab:spca_comp_times} reports timings for a fixed sample size \code{n = 2000} and an increasing number of variables. 
\begin{table}[H]
\centering
\begin{minipage}{0.95\linewidth}
\begin{verbatim}
  dimension min time median time mem_alloc
1   2kx100   12.45ms      14.4ms    1.69MB
2   2kx300  182.18ms     191.1ms    5.39MB
3   2kx500  783.53ms     823.1ms    9.69MB
4   2kx1000    5.46s        5.8s   23.12MB

\end{verbatim}
\end{minipage}
\caption{spca computational times and memory allocation for matrices of increasing size. '2k' stands for 2000.}\label{tab:spca_comp_times}
\end{table}

The median runtime grows from approximately 14 ms at $p = 100$ to 5.8 s at $p = 1000$; the implied scaling exponent rises approximately from 2.4 to 2.8 across the grid, reflecting the transition from the $O(np^2)$
covariance computation to the $O(p^3)$ eigendecomposition and least-squares operations as $p$ increases. 
Memory allocation grows essentially linearly in $p$ and remains below 25 MB throughout, indicating that runtime rather than storage is the limiting factor in high dimensions.

\subsection{Comparison with conventional SPCA}
Scalability experiments were conducted to compare \pkg{spca} with \pkg{abess} version 1.3 \citep{abess}, a recent package featuring a dedicated optimized \proglang{C++} backend that explicitly targets computational efficiency, making it a  challenging benchmark. This package was preferred to \pkg{elasticnet} because, as shown in Section~\ref{sec:comparisons}, it is significantly slower even at moderate dimensions and including it in a scalability study would add little insight.

We first ran \fct{spca} with default settings on each target matrix to obtain the cardinalities of the LS-SPCA solutions. We then benchmarked up to 100 iterations of \fct{spca}, both with the full eigendecomposition and with the power method enabled (\code{ls-spca-PM}), and compared them with \pkg{abess}'s \fct{abesspca} function. The \fct{abesspca} call used the same number of components and the same cardinalities as the LS-SPCA solution, with the following command:
\begin{Code}
    abesspca(X = X,
             type = "predictor",
             sparse.type = "kpc",
             kpc.num = r,
             support.size = as.list(ls_spca$cardinality
             )
\end{Code}
The tuning parameters \code{tune.type} and \code{tune.path} are irrelevant in this comparison because \code{support.size} is passed explicitly.

We compared \fct{spca}, with and without the power method, against \fct{abesspca} on two tall matrices of size $1000\times 100$ and $3000\times 1000$. The \fct{abesspca} solutions were computed using the same cardinalities selected by \fct{spca}. The results are presented in Table~\ref{tab:comp_lsspca}.

\begin{table}[H]
\centering
\begin{minipage}{0.95\linewidth}
\begin{verbatim}
      dimension    function    min time median time   mem_alloc
1 1,000 x 100    ls_spca         0.02ms      0.02ms       2.6MB
2 1,000 x 100    ls_spca_PM      0.01ms      0.01ms       2.6MB
3 1,000 x 100    ab_spca         0.14ms      0.14ms      86.2MB
4 3,000 x 1,000  ls-spca         7.04ms      7.23ms        49MB
5 3,000 x 1,000  ls-spca PM      1.03ms      1.05ms      48.4MB
6 3,000 x 1,000  ab_spca         18.2ms     18.44ms     325.8MB

\end{verbatim}
\end{minipage}
\caption{Computation times and memory allocation for default spca(), spca() with
  the power method, and abesspca(). The abesspca() fits use the same
  component  cardinalities selected by spca().}\label{tab:comp_lsspca}
\end{table}

The results show that both the default \fct{spca} and the power-method version are faster than \fct{abesspca}. On the smaller matrix, the \fct{spca} fits are approximately 7 and 14 times faster, respectively. For the larger matrix, these factors decrease to about 2.5 for the default version but increase to about 18 for the power-method version. These comparisons are conservative for \pkg{spca}, because \fct{abesspca} was run with fixed cardinalities rather than allowing for selection.
\subsubsection{Fat matrices}
Table~\ref{tab:comp_lsspcaFat} shows the performances of the three functions tested in Section~\ref{sec:lm_vs_ab_tall}  on a $100\times 1000$ fat matrix.
\begin{table}[H]
\centering
\begin{minipage}{0.95\linewidth}
\begin{verbatim}
    function min time median time mem_alloc
1 ls-spca      23.9ms      24.7ms   160.1KB
2 ls-spca-PM   18.7ms        20ms   159.2KB
3 ab-spca     627.2ms     799.6ms    88.9MB

\end{verbatim}
\end{minipage}
\caption{Computation times and memory allocation for default spca(), spca() with the power method, and abesspca() applied to a fat matrix. The abesspca() fits use the same component cardinalities selected by spca().}\label{tab:comp_lsspcaFat}
\end{table}

The difference in efficiency is even larger for the fat matrix. Both \fct{spca} variants complete in approximately 20--25 ms and allocate very little memory, whereas \fct{abesspca} takes approximately 0.8 s and allocates almost 90 MB. This suggests that the dedicated fat-matrix backend in \pkg{spca} is effective. This comparison may also reflect that \pkg{abess} is not specifically optimized for fat matrices.
\section{Concluding remarks}\label{conc}
The \pkg{spca} package implements LS-SPCA as a practical tool for replacing ordinary PCs with sparse, more interpretable components. The examples show that LS-SPCA solutions can remain very close to the original PCs while using substantially fewer variables. At the same time, the package gives users considerable flexibility to choose among computational methods, variable selection strategies, stopping rules, and target levels of approximation.

The package also provides methods and helper functions for evaluating different fits numerically and visually, including comparisons with standard PCA. The \code{compare_spca} helper compares two or more SPCA solutions, whereas \code{new_spca} allows users to bring in SPCA solutions computed outside \pkg{spca} and evaluate them within the same framework.

The comparisons presented further clarify the difference between LS-SPCA and conventional SPCA. While conventional SPCA can produce useful sparse components, these components tend to deviate more from the corresponding PCs and are more mutually correlated. In this sense, they are closer in spirit to oblique rotations used in factor analysis than to sparse approximations of PCA.

In the future the package will be developed to prioritize scaling the eigendecomposition solver to accommodate larger values of $p$, potentially through the implementation of truncated methods. Additionally, support for grouped and structured sparsity patterns, using prior knowledge to guide variable and component selection may be added.

In conclusion, the \pkg{spca} package is not solely dedicated to fitting the LS-SPCA model. It offers a fast \proglang{C++} backend, specialized algorithms for tall and fat matrices, and an \proglang{R} interface designed to inspect and compare the solutions. The fitted \class{spca} objects provide essential metrics for evaluating a sparse PCA solution, including cardinality, VEXP, CVEXP, correlation with the PCs, and correlations among sPCs. Together with graphical and numerical summaries and comparison methods,  \pkg{spca} provides a flexible and efficient environment for LS-SPCA analyses that can either replace or work alongside PCA.

\bibliography{JSS}
\end{document}